\documentclass[12pt]{article}
\usepackage{amsfonts}
\usepackage{epsfig}
\usepackage{graphicx}
\usepackage{amsmath}
\begin{document}

\title{Casimir-Polder interaction in the presence of parallel walls}
\author{F C Santos$^\dagger$, J. J. Passos Sobrinho and A. C. Tort$^\ddagger$ \\
Instituto de F\'{\i}sica\break \\
Universidade Federal do Rio de Janeiro\break \\
Cidade Universit\'aria - Ilha do Fund\~ao - Caixa Postal 68528\\
21945-970 Rio de Janeiro RJ, Brasil.}
\date{\today}
\maketitle

\begin{abstract}
Making use of the quantum correlators associated with the Maxwell field vacuum
distorted by the presence of plane parallel material surfaces we derive the Casimir-Polder in the presence of plane parallel conducting walls and in the presence of a conducting wall and a magnetically permeable one.
\end{abstract}

\noindent PACS: 11. 10. -z; 12. 20. -m

\vfill

\noindent $^{\dagger }$ {e-mail: filadelf@if.ufrj.br}

\noindent $^{\ddagger }$ {e-mail:tort@if.ufrj.br}\newpage

In 1948, Casimir and Polder \cite{CasimirPolder48} taking into account a suggestion made by experimentalists evaluated the interaction potential between two eletrical polarizable molecules separated by a distance $r$ including the effects due to the finiteness of the speed of propagation of the electromagnetic interaction, i.e.: of the retardment. Casimir and Polder
showed that the retardment causes the interaction potential to change from a $r^{-6}$ power law to a  $r^{-7}$ power law. In the same paper, Casimir and Polder also analyzed the retarded interaction between an atom and a conducting wall and showed that the interaction potential in this case varies according to a  $r^{-4}
$ power law,  where now  $r$ is the distance between the atom and the wall. For an introduction on these subjects see \cite{Milonni}. Here we wish to show how it is possible with the help of the so called renormalized electromagnetic field correlators, in our case the ones that take into account the presence of the boundary conditions imposed on the fields, to reobtain the piece of Casimir
and Polder's result for the atom-wall interaction that depends on the
distortion of the vacuum oscillations of the electromagnetic field caused by the presence of parallel walls. The electromagnetic field correlators for the case of two parallel perfectly conducting surfaces separated by a distance $a$ were evaluated in \cite{Lü&Ravndal85} and in \cite{JPA1999}. For the case of a perfectly conducting plane wall and a perfectly permeable plane wall, a setup first introduced by Boyer \cite{Boyer74}, they were calculated in \cite{JPA1999}. These mathematical objects, closely related to the pertinent electromagnetic Green's functions, were also employed to obtain an alternative view of the Casimir effect \cite{Casimir} through the quantum version of the Lorentz force between the walls \cite{EJP2003}.

Let us first recall some aspects concerning electrically and magnetically polarizable bodies \cite{Jackson3rd}. From a classical point of view the induced eletrical polarization density $\mathbf{P}$ can be thought of as a function of the electric and magnetic
fields $\mathbf{E}$ and $\mathbf{B}$. In many cases only the dependence on the eletric field is relevant. It can be shown that under conditions for which the effects of the retardment (i.e., of the finiteness of the speed of light) must be taken into account it suffices to consider the static
eletrical polarizability $\alpha \left( 0\right) $ only, see for instance \cite{Milonni} and references therein. If the electric field changes by $\delta \mathbf{E} $, the interaction between the polarizable body and the electric field will change according to $\delta V=-\mathbf{P}\left[ \mathbf{E}\right]
\cdot \delta \mathbf{E}=-\alpha \left( 0\right) \mathbf{E\cdot }\delta
\mathbf{E}$. Therefore, if the field changes from zero to a finite value $\mathbf{E}$, the interaction energy is $V_{E}=-\alpha \left( 0\right)
\mathbf{E}^{2}/2$. In the quantum version of this interaction potential we
must replace  $\mathbf{E}^{2}$ by its vacuum expectation value, $%
\left\langle \mathbf{\hat{E}}^{2}\right\rangle _{0}$. The same arguments hold when we consider the magnetization  $\mathbf{M}$. The quantum interaction potential between a magnetically polarizable atom and the magnetic field is given by $V_{M}=-\beta \left( 0\right) \left\langle\mathbf{B}^{2}/2\right\rangle_{0}$, where $\beta \left(
0\right) $ is the static magnetic polarizability. In order to proceed we must know the vacuum expectation values of the quantum field operators $\mathbf{E}^2$ and $\mathbf{B}^2$. This means to evaluate explicitly the vacuum expectation values of the so called electromagnetic field correlators  $E_i\left(\mathbf{r},t\right)E_j\left(\mathbf{r},t\right)$, $B_i\left(\mathbf{r},t\right)B_j\left(\mathbf{r},t\right)$, and $E_i\left(\mathbf{r},t\right)B_j\left(\mathbf{r},t\right)$, in the presence of external conditions, i.e., boundary conditions. A regularization recipe will also be necessary. Fortunately these objects were calculated before and we can limit ourselves to make use of the results. 

For the case of two parallel conducting walls separated by a fixed distance $a$ we have \cite{Lü&Ravndal85,JPA1999}
\begin{eqnarray}
\langle E_{i}(\mathbf{r},t)E_{j}(\mathbf{r},t)\rangle _{0}
&=&\left( {\frac{\pi }{a}}\right) ^{4}{\frac{2}{3\pi }}\left[ \left( -\delta
^{\Vert }+\delta ^{\bot }\right) _{ij}\;{\frac{1}{120}}+\delta _{ij}F(\xi )%
\right] \,,  \label{ECORRCASIMIR}
\end{eqnarray}
where $\delta _{ij}^{\Vert }:=\delta _{ix}\delta _{jx}+\delta _{iy}\delta
_{jy}$ and $\delta _{ij}^{\bot }:=\delta _{iz}\delta _{jz}$. The function $\,F\left( \xi \right) $ with $\xi:=\pi z/a$ is defined by
\begin{equation}
F\left( \xi \right) :=-\frac{1}{8}\frac{d^{3}\,}{d\xi ^{3}}\frac{1}{2}\cot
\left( \xi \right),
\end{equation}
and its expansion about $\xi =0$ is given by
\begin{equation}\label{Fapp}
F\left( \xi \right) \approx \frac{3}{8}\xi^{-4}+\frac{1}{120}+O\left(\xi^2\right).
\end{equation}
Near $\xi =\pi $ (which corresponds to $z=a$) we make the
replacement $\xi \rightarrow \xi -\pi $. Notice that due to the behavior of $%
F\left( \xi \right) $ near $\xi =0,\pi$, divergences control the
behavior of the correlators near the plates.

The magnetic field correlators are \cite{Lü&Ravndal85, JPA1999}
\begin{equation}
\langle B_{i}(\mathbf{r},t)B_{j}(\mathbf{r},t)\rangle _{0}=\left( {\frac{\pi
}{a}}\right) ^{4}{\frac{2}{3\pi }}\left[ \left( -\delta ^{\Vert }+\delta
^{\bot }\right) _{ij}\;{\frac{1}{120}}-\delta _{ij}F(\xi )\right] \,.
\label{BCORRCASIMIR}
\end{equation}
A direct evaluation shows that the correlators $<E_{i}(\mathbf{r},t)B_{j}(\mathbf{r},t)\rangle _{0}$ are zero.

For the case of perfectly conducting plane wall and a magnetically permeable one results are \cite{JPA1999}
\begin{equation}
\left\langle \hat{E}_{i}\left( \mathbf{r},t\right) \hat{E}_{j}\left( \mathbf{%
r},t\right) \right\rangle _{0}=\left( \frac{\pi }{a}\right) ^{4}\frac{2}{%
3\pi }\left[ \left( -\frac{7}{8}\right) \frac{\left( -\delta ^{\Vert
}+\delta ^{\perp }\right) _{ij}}{120}+\delta _{ij}\,G\left( \xi \right) %
\right],   \label{EcorrBoyer}
\end{equation}
and
\begin{equation}
\left\langle \hat{B}_{i}\left( \mathbf{r},t\right) \hat{B}_{j}\left( \mathbf{%
r},t\right) \right\rangle _{0}=\left( \frac{\pi }{a}\right) ^{4}\frac{2}{%
3\pi }\left[ \left( -\frac{7}{8}\right) \frac{\left( -\delta ^{\Vert
}+\delta ^{\perp }\right) _{ij}}{120}-\delta _{ij}\,G\left( \xi \right) %
\right].   \label{BcorrBoyer}
\end{equation}
Observe that near $\xi =0$ the function $G\left( \xi \right) $ behaves as
\begin{equation}
G\left( \xi \right) =\frac{3}{8}\xi ^{-4}-\frac{7}{8}\,\frac{1}{120}+O\left(
\xi ^{2}\right) ,  \label{Gapp1}
\end{equation}
near $\xi =\pi$, however, its behavior is slightly different
\begin{equation}
G\left( \xi \right) =-\frac{3}{8}\left( \xi -\pi \right) ^{-4}+\frac{7}{8}\,%
\frac{1}{120}+O\left[ \left( \xi -\pi \right) ^{2}\right] .  \label{Gapp2}
\end{equation}
Again, a direct calculation shows that $\left\langle \hat{E}_{i}\left(
\mathbf{r},t\right) \hat{B}_{j}\left( \mathbf{r},t\right) \right\rangle
_{0}=0$ for this case also. As before the divergent behavior of the correlators near the plates we are
interested in is an effect of the distortions of the electromagnetic
oscillations with respect to a situation where the plates are not present. The correlators given by (\ref{ECORRCASIMIR}), (\ref{BCORRCASIMIR}), (\ref{EcorrBoyer}), and (\ref{BcorrBoyer})
allow us to obtain in a straightforward way expressions for the interaction
potential energy between an electrically or magnetically polarizable atom placed between the walls and the walls.

Let us consider first the case of an electrically polarizable atom or molecule placed between two perfectly conducting parallel walls. Suppose that the atom is placed at a distance $%
z$ from the conducting wall placed at $z=0$. The interaction potential
between the atom and the walls is given by
\begin{equation}
V_{E}\left( z\right) =-\frac{1}{2}\alpha \left( 0\right) \left\langle
\mathbf{\hat{E}}^{2}\left( z\right) \right\rangle _{0},
\end{equation}
where $\alpha \left( 0\right) $ is the static polarizability of the
molecule. Making use of (\ref{ECORRCASIMIR}) we can
evaluate $\left\langle \mathbf{\hat{E}}^{2}\left( z\right) \right\rangle _{0}
$ and using the above equation we obtain
\begin{equation}
V_{E}\left( z\right) =-\frac{\alpha \left( 0\right) \pi ^{3}}{3a^{4}}\left[
3F\left( \frac{\pi z}{a}\right) -\frac{1}{120}\right].
\end{equation}
Making use of (\ref{Fapp}) and taking the limit $a\rightarrow \infty $ we obtain the single wall limit of the interaction potential between an electrically polarizable atom and a conducting
wall,
\begin{equation}
V_{E}\left( z\right) =-\frac{3\alpha \left( 0\right) }{8\pi z^{4}},
\end{equation}
in agreement with \cite{Casimir49,Boyer69}; see also \cite{Milonni}. Consider now a magnetically polarizable atom or molecule placed between the two conducting walls. The
interaction potential in this case will be given by
\begin{equation}
V_{M}\left( z\right) =+\frac{\beta \left( 0\right) \pi ^{3}}{3a^{4}}\left[
3F\left( \frac{\pi z}{a}\right) +\frac{1}{120}\right] ,
\end{equation}
where we made use of (\ref{BCORRCASIMIR}). If the atom or molecule is simultaneously electrically and magnetically
polarizable the interaction potential will be simply $V\left( z\right)
=V_{E}\left( z\right) +V_{M}\left( z\right) $, that is
\begin{equation}
V\left( z\right) =-\left( \alpha \left( 0\right) -\beta \left( 0\right)
\right) \frac{\pi ^{3}}{a^{4}}F\left( \frac{\pi z}{a}\right) +\left( \alpha
\left( 0\right) +\beta \left( 0\right) \right) \frac{\pi ^{3}}{360a^{4}}.
\label{V}
\end{equation}
The single conducting wall limit $(a\rightarrow \infty )$ of (\ref{V}) is
easily obtained with the help of (\ref{Fapp}). The result is:
\begin{equation}
V\left( z\right) \approx -\frac{3}{8\pi z^{4}}\left( \alpha \left( 0\right)
-\beta \left( 0\right) \right) ,  \label{Vapp}
\end{equation}
which is in agreement with \cite{Casimir49,Boyer69}.

The polarizable atom or molecule can be also placed between a conducting
plate at $z=0$ and a permeable one at $z=a$. In this case, making use of (%
\ref{EcorrBoyer}) e (\ref{BcorrBoyer}) a straightforward calculation leads
to the following result
\begin{equation}
V\left( z\right) =-\left( \alpha \left( 0\right) -\beta \left( 0\right)
\right) \frac{\pi ^{3}}{a^{4}}G\left( \frac{\pi z}{a}\right) +\left( \alpha
\left( 0\right) +\beta \left( 0\right) \right) \left( -\frac{7}{8}\right)
\frac{\pi ^{3}}{360a^{4}}.
\end{equation}
There are now two single walls limits to be considered. Near the conducting
plate at $z=0$ the potential is given by (\ref{Vapp}), but near the
perfectly permeable plate at $z=a$, the potential is repulsive and given by
\begin{equation}
V\left( z\right) \approx +\frac{3}{8\pi \left( z-a\right) ^{4}}\left( \alpha
\left( 0\right) -\beta \left( 0\right) \right) ,
\end{equation}
where we made use of (\ref{Gapp2}). These last results are to our knowledge new.  

It is important to keep in mind that, as mentioned before,
we have dealt with a part of the interaction between an atom and two or one
walls. The contribution of the interaction between the electric/magnetic dipole moment and its images was utterly neglected. Therefore, the results refer only to the contribution of the quantum vacuum distorted by one or two walls
to the total interaction potential. With this proviso we can state that the Casimir-Polder interaction shows certain aspects of the quantum structure of the vacuum confined between the plane surfaces in question. The atom acts as a probe of the confined quantum vacuum, particularly near the walls.

\end{document}